\documentstyle[aps,pra,twocolumn,psfig]{revtex}

\begin{document}

\title{Cold atoms: A new medium for quantum optics}
\author{A. Lambrecht, J.M. Courty, S. Reynaud, E. Giacobino \\
Laboratoire Kastler Brossel, UPMC, ENS, CNRS\\
Universit\'{e} Pierre et Marie Curie, 4 place Jussieu, F 75252 PARIS}
\date{{\sc Applied Physics} {\bf B60}, p.129 (1995)}
\maketitle

\begin{abstract}
{\bf Abstract.} Laser cooled and trapped cesium atoms have been used as a 
non-linear medium
in a nearly resonant cavity. A study of the semi-classical dynamics of the
system was performed, showing bistability and instabilities. In the quantum
domain, squeezing in a probe beam having interacted with this system was
demonstrated.\\[2mm]
PACS: 32.80.Bx; 32.80.Pj; 42.50.Dv; 42.50.Lc; 42.65.Pc
\end{abstract}

\vspace{5mm}

Cold atoms in a magneto-optical trap constitute a quite new non-linear
material, the properties of which are far from being fully investigated.
With the very low temperatures that can be achieved, the atoms can be
considered as virtually motionless and the Doppler width of their resonances
is smaller than the natural linewidth. In the vicinity of resonance lines,
one can expect large non-linear dispersive effects associated with very
little absorption, but it is only recently that a cloud of cold atoms was
placed in a resonant optical cavity to study its semi-classical dynamics \cite{Giac93}. 
As we show in the first part of this paper, cold
cesium atoms in a cavity have a very rich non-linear dynamics, exhibiting
not only bistability but also instabilities. Analysing the non-linear
dynamics of laser cooled atoms in a cavity was a prerequisite to the
investigation of quantum noise in order to find the appropriate conditions
to study the quantum properties of a light beam going out of the cavity.

The generation of quadrature squeezed light through interaction with atomic
media has been the subject of numerous theoretical studies and looked
particularly promising at the condition that there is no Doppler effect. All
the experiments up to date were performed on atomic beams. In the second
part of this paper, we report on the quantum properties of cold atoms in a
cavity and show that it is possible to detect quadrature squeezing in a
probe beam that has passed through the cavity.

\section{Dynamics of cold atoms in a cavity}

We investigate the dynamical non-linear behaviour of an optical cavity
containing cold cesium atoms by sending a circularly polarisedprobe beam
into the cavity. The atoms are cooled and trapped in a magneto-optical trap,
operating in the standard way \cite{Monroe90} with three orthogonal
circularly polarised trapping beams generated by a Ti:Sapphire laser and an
inhomogeneous magnetic field.

The cavity is a 25cm long linear asymmetrical cavity, close to
half-confocal, with a waist of 260$\mu $m, built around the cell (Fig. 1).
Losses due to the two AR coated windows are of the order of 1\%. The input
mirror has a transmission coefficient of 10\%, the end mirror is highly
reflecting, which gives a cavity linewidth of 5MHz. The cavity is in the
symmetry plane of the trap, making a 45$^{\circ }$ angle with the two
trapping beams that propagate in this plane.
\begin{figure}
\centerline{\psfig{figure=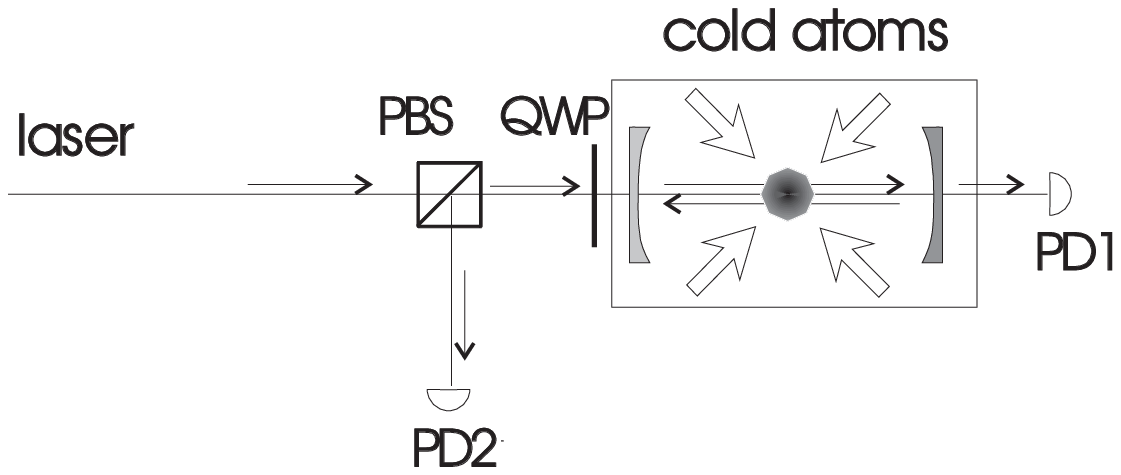,width=7cm}}
\caption{Experimental set-up showing the cell containing the cold atom cloud in
an optical cavity; PBS: polarising beamsplitter, QWP: quarter-wave plate,
PD1, PD2: photodiodes; PD1 and PD2 measure the powers respectively
transmitted and reflected by the cavity.}
\end{figure}

The trapping beams are detuned by 2.5 times the linewidth of the upper state
($\Gamma/2\pi=5.2$MHz) on the low frequency side of the $6S_{1/2} F=4$ to $6P_{3/2} F=5$ transition, yielding a cloud of cesium atoms the typical
temperature of which is of the order of 1 mK. The diameter (2.5cm) and
power (20mW/cm$^2$) of the trapping to obtain large clouds (about 8mm in
diameter) with densities of the order of beams allow us 10$^{10}$~atoms/cm$%
^{3}$. The number of trapped atoms located in the probe beam, which is the
relevant parameter in the experiment, measured from the change in the
transmission of the cavity in the presence of atoms, is found to be ranging
between 10$^{7}$ and 10$^{8}$ depending on the pressure of the background
gas. As usual, the atoms non resonantly excited into the $6P_{3/2}F=4$ state
and falling back into the $F=3$ ground state are repumped into the cooling
cycle by a laser diode tuned to the $6S_{1/2} F=3$ to $6P_{3/2} F=4$
transition.

The probe beam, generated by the same laser, can be detuned by 0 to 130MHz
(corresponding to 0-25$\Gamma $) on either side of the $6S_{1/2} F=4$ to 
$6P_{3/2} F=5$ atomic transition. We measure the power of the probe beam
transmitted through the cavity while scanning the cavity length for a fixed
value of the input intensity as shown in Fig. 2, or the input intensity for a
fixed value of the detuning. The recording shows the characteristic
hysteresis cycle due to bistability, where the output power switches
abruptly between low and high values when the length of the cavity is
scanned.
\begin{figure}
\centerline{\psfig{figure=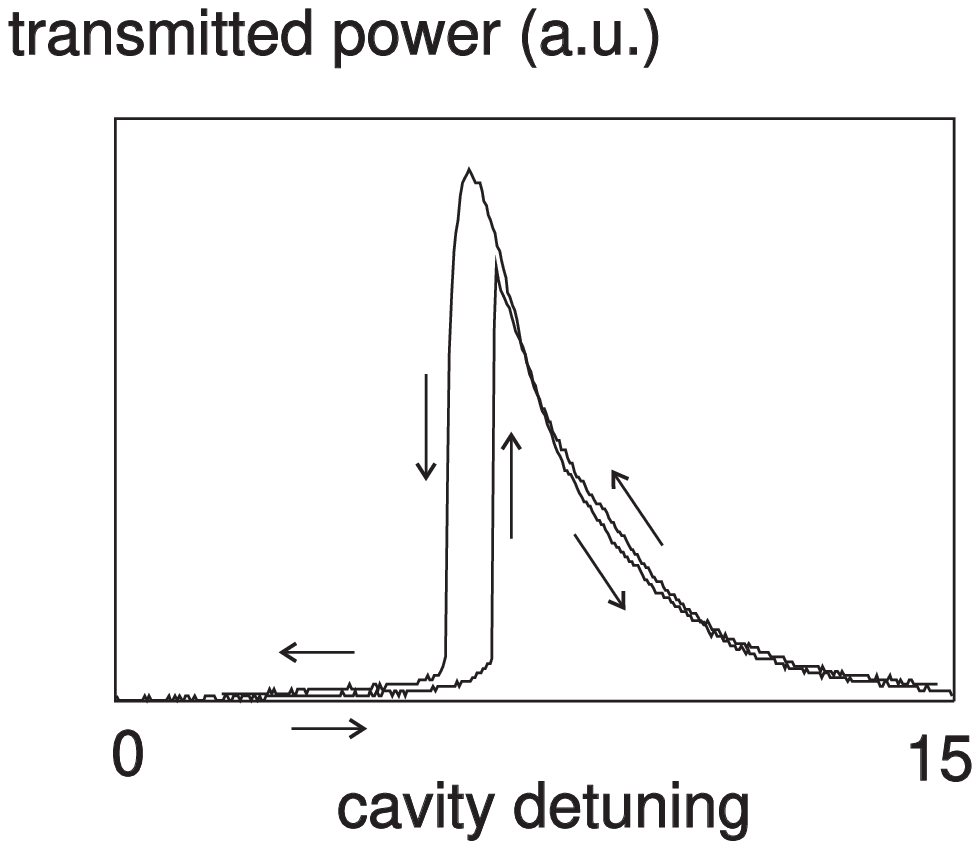,width=5.5cm}}
\caption{Recording showing the bistable switching from low to high transmission
and back when the cavity length is scanned across resonance. The trapping
beams are on. The probe laser is detuned by 22$\Gamma $ on the high
frequency side of the atomic transition; the input power is 100$\mu$W.
Cavity detuning is in units of the cavity linewidth $\gamma_{\rm 
cav}/2\pi\tau$.}
\end{figure}

The shape of the curves is compatible with the theory of bistability with
two-level atoms. However, the $F=4$ to $F=5$ transition under investigation
is far from being a two-level one. In the presence of the trapping beams, it
can be approximately described as a set of two-level systems: the circularly
polarised probe beam interacts with all $m_{F}$ to $m_{F+1}$ transitions from
the ground to excited state, while the trapping beams randomise the ground
state population among the various Zeeman sublevels. In this sense the
bistability curve shown in Fig. 2 has to be understood as an effective
bistability, in which all sublevels participate.

When the probe beam is strong enough, the induced optical pumping cannot be
neglected. To thoroughly investigate the dynamics of the system, it was
better to avoid a competition between trapping beams and probe beam in the
interaction with the atoms. Consequently, the trapping laser beams are
turned off after the trap has been loaded. The time left for the measurement
is about 20ms before most of the atoms have escaped out of the interaction
region due to free fall and expansion of the cloud. A significant measure of
the number of atoms consists in the bistability parameter $C$, which is the
ratio of linear absorption of the atoms at resonance to the energy
transmission coefficient of the cavity. Right after the atoms have been
released, the bistability parameter can be as high as 400.

In a broad range of experimental parameters, not only bistability but
instabilities are observed. Figure 3 shows a recording of the transmitted
power of the cavity for two different values of the input power, while the
cavity length is scanned. At low input powers of the order of $50\mu $W
oscillations appear on the left hand side of the cavity resonance. This side
is the one on which bistable switching would occur in a saturated two-level
atomic system. In particular it has to be noted that the first trace (input
power $100\mu $W) was recorded under the same experimental conditions as
the trace in Fig. 2, except that the trapping beams were switched off. While
in presence of the beams the system is clearly bistable, in their absence
instabilities appear instead. At high input powers (about $300\mu$W) the
oscillations disappear and only bistability persists.
\begin{figure}
\centerline{\psfig{figure=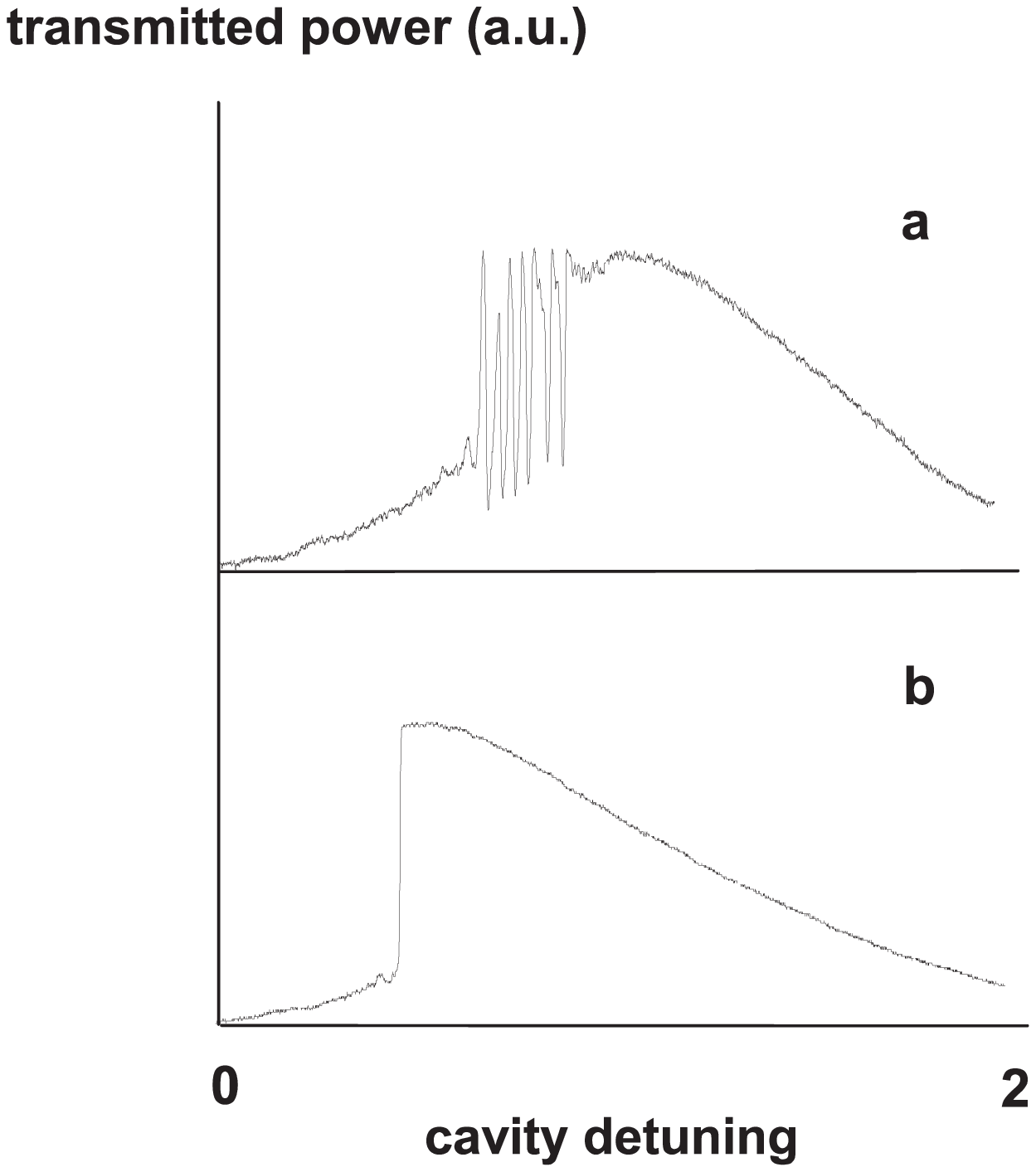,width=5cm}}
\caption{Recording of the power transmitted through the cavity containing cold
atoms while the cavity length is scanned across resonance. Input powers are 
$P_{{\rm in}}=100\mu$W (a) and P$_{{\rm in}}$=350$\mu$W (b)
respectively. The atomic detuning is 22$\Gamma$ on the blue side of the atomic
transition, the trapping beams are off. Cavity detuning is in units of the
cavity linewidth $\gamma_{\rm cav}/2\pi\tau$.}
\end{figure}

Instabilities are present within the whole range of accessible detunings on
either side of the resonance. Their frequency is comprised between 100kHz
and a few MHz. They can also be observed when the cavity length is scanned
by the escape of the atoms from the original cloud, which changes the index
of refraction.

To understand these oscillations, the hyperfine and Zeeman structure of the
considered states and the various possible optical pumping effects have to
be taken into account. Instabilities \cite{Gius87,Penna93,Mitsch83} linked to optical pumping were predicted and observed in alkali
vapours. However, this system is rather intricate to describe, due to the
interaction of the field with atoms in various velocity classes and
hyperfine sublevels.

The situation is much simpler with cold atoms, where one can consider that
the field interacts with one hyperfine transition only. Owing to this, we
have been able to understand the origin of the observed oscillations with a
simple model. Roughly, as shown below, they result from the competition
between two non-linear processes, saturation of the optical transition,
which tends to decrease the linear index of refraction of the atomic medium
when the intensity increases, and optical pumping that increases the index
of refraction by accumulating the atoms in magnetic sublevels with a high 
$m_{F}$ number, which have the largest coupling coefficient with the light.

While the saturation of the optical transition is virtually instantaneous,
it takes a comparably long time to complete the optical pumping to the
highest mF sublevels of the ground and excited states, starting from an
equally distributed population in the ground state. Even for intensities
close to or larger than the saturation intensity the optical pumping rate is
much smaller than the natural linewidth \cite{Giac94}. The
relaxation oscillations are a consequence of the significant difference in
the characteristic times of these processes.

Although the behaviour of the system involving many variables (all hyperfine
Zeeman populations and coherences and the field) is quite complex, the
underlying mechanism can be explained with a simple model, based on
essentially two differential equations, one for the evolution of the
intracavity field and the other for the atomic orientation in the ground
state.

The change of the intracavity field a on a round trip of time duration $\tau$ 
is due to the driving field $\alpha _{{\rm in}}$ entering through the
coupling mirror of transmission t, to the mirror decay coefficient gcav
(with $\gamma _{{\rm cav}}$=$t^{2}/2$) and to the round trip phase shift 
$\phi _{{\rm cav}}$:
\begin{equation}
\tau \frac{{\rm d}\alpha }{{\rm d}t}=t \alpha_{{\rm in}}-(\gamma _{{\rm %
cav}}-i\phi _{{\rm cav}})\alpha.   \label{champ}
\end{equation}
The total phase shift in the cavity $\phi _{{\rm cav}}$ writes:
\begin{equation}
\phi _{{\rm cav}}=\phi _{0}+\phi _{{\rm L}}(1+p)+\phi _{{\rm NL}},
\end{equation}
where $\phi _{0}$ is the round trip phase shift in the empty cavity, $\phi
_{L}$ is the term due to the linear index of the atoms and $p\phi _{L}$ the
additional contribution associated with the ground state orientation of the
atoms induced by optical pumping; $\phi _{NL}$ is the Kerr-like phase shift
due to saturation of two-level atoms.

The ground state orientation p increases with the intracavity intensity $I$
at a rate $\beta I$ and decays at a rate $\gamma _p$by magnetic
precession in transverse fields and by transitions to other hyperfine
sublevels (via non resonant transitions):
\begin{equation}
\frac{{\rm d}p}{{\rm d}t}=-\gamma _{{\rm p}}p+\beta I(1-p).  \label{pump}
\end{equation}
In the presence of optical pumping equations (\ref{champ}) and (\ref{pump}) have
to be solved numerically. In some range of initial conditions, oscillations
and limit cycles are found in the intracavity intensity as well as in the
output intensity. The calculated output intensity as a function of the
cavity length is shown in Fig. 4 for both a low and a high value of the input
intensity. It can be seen that the curves reproduce the experimental
recordings in a satisfactory way. Oscillations are essentially manifest when
the different time constants involved are such, that (already small)
variations in the atomic ground state orientation scan the cavity back and
forth around the bistable switching point and therefore lead to periodic
abrupt changes in the intracavity power. Only at high input powers
instabilities disappear, as the optical pumping is fast enough to bring all
atoms in the sublevel with the highest magnetic number before the
oscillations can start. In this sense also the absence of the instabilities
in presence of the trapping beams (Fig. 2) can now be understood, because
optical pumping by the probe beam is counteracted by the trapping beams,
which themselves produce a strong optical pumping of the atoms, which tends
to randomise the populations in the various sublevels. 
\begin{figure}
\centerline{\psfig{figure=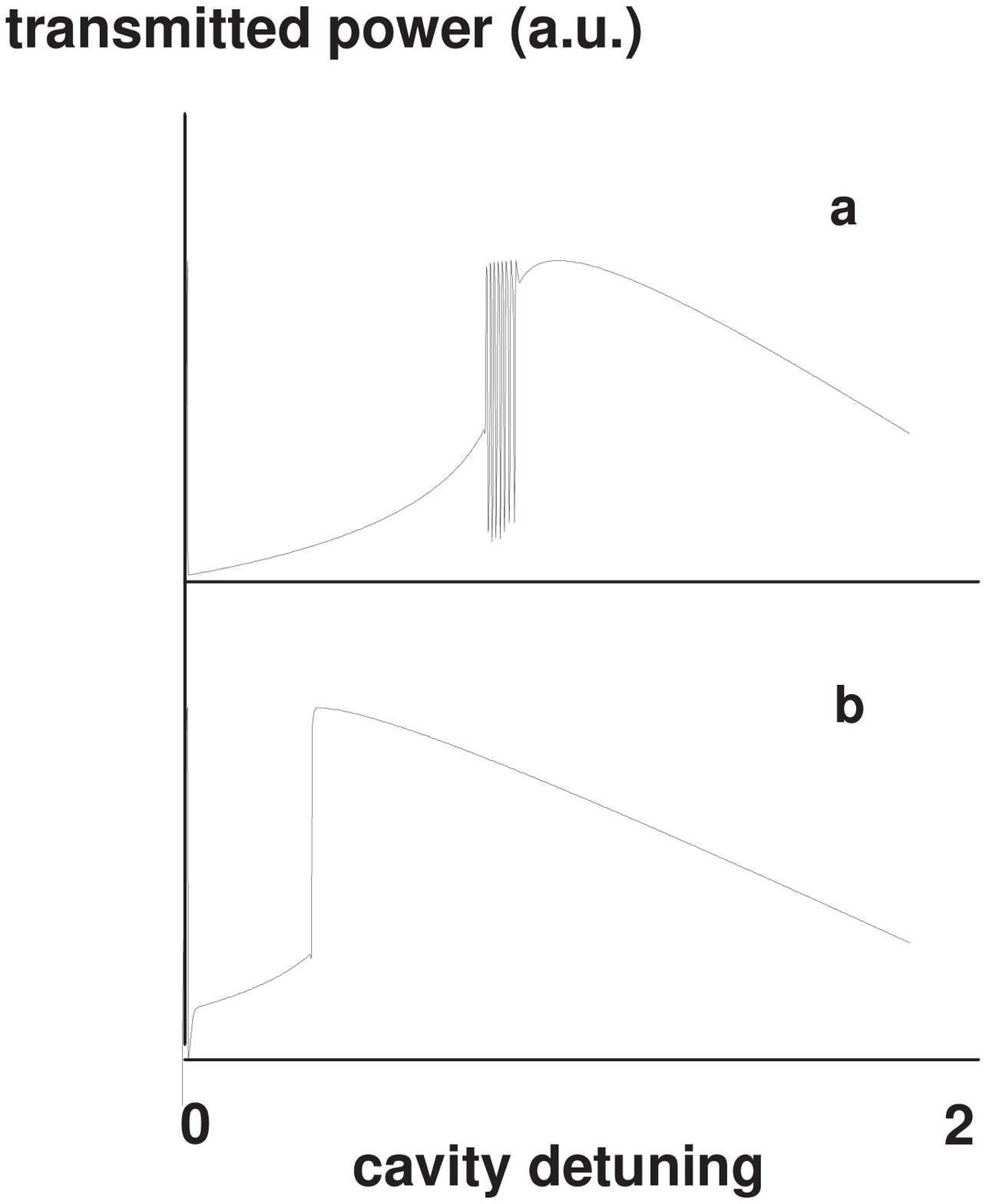,width=4.6cm}}
\caption{Calculated power transmitted through the cavity containing cold atoms
as a function of cavity detuning (normalised to the cavity linewidth $\gamma
_{{\rm cav}}$/2$\pi \tau $). Input powers correspond respectively to 1.3 (a)
and 2 (b) times the bistability threshold power in the absence of optical
pumping.}
\end{figure}

The validity of the model was checked by further experiments. When the atoms
are released from the trap, it is possible to optically pump them into the 
$m_{F}=4$ magnetic sublevel of the ground state with an additional
circularly polarised beam parallel to the probe beam, but closer to the
atomic resonance. This prepumping is done in the presence of a magnetic
field directed along the cavity. In such conditions the instabilities
disappear. If the magnetic field is absent, the orientation created by the
pump field is destroyed by the Larmor precession in transverse magnetic
fields and the instabilities persist.

A more complete theoretical treatment of the atomic non-linearity and of the
optical pumping was also performed, where absorption and saturation of the
optical non-linearity were taken into account. The general behaviour remains
the same as the one predicted by the simple model. 

If cold atoms are to be used as a non-linear medium to produce squeezed
light, clearly instabilities have to be avoided. The results presented so
far show that, although the behaviour of the system is more complicated than
anticipated, it is possible to find conditions where cold cesium atoms
behave as two-level atoms.

\section{Quantum noise measurements}

Nearly perfect squeezing is expected from a cavity containing a pure Kerr
medium \cite{Lugia92,Collett85,Reynaud89}. When motionless
two-level atoms are used, the quantum noise reduction is predicted to
persist in spite of added fluorescence noise. Specific theories have to be
used to take these effects into account \cite{Orozco87,Castelli88,Reid88,Hilico92}. Squeezed vacuum was first observed in a
four-wave mixing experiment using an atomic beam as a non-linear medium \cite
{Slusher85}. Later, squeezing in a beam of finite mean intensity (bright
squeezing) was demonstrated at the output of a bistable cavity traversed by
an atomic beam~\cite{Raizen87,Hope92}.

Cold atoms seem to be good candidates to achieve quantum noise reduction,
since their motion is much better controlled than in an atomic beam.
However, observation of squeezing requires that the trapping beams are
absent. Otherwise a part of the atoms would be constantly excited to the $%
6P_{3/2}$ state and fluoresce back to the ground state, causing excess
noise. This condition imposed the study of the non-linear dynamics of the
cavity in absence of the trapping beams. To avoid excess noise and
instabilities both trapping beams and magnetic field gradient are therefore
turned off during the noise measurement; residual transverse fields are
largely compensated by a homogeneous magnetic field.

The experimental set-up used to study the quantum fluctuations of a probe
beam after interaction with the system is shown in Fig. 5. The linear cavity
containing the atom cloud was described in section 1. It has only one
coupling mirror of 10\% transmission, so as to be close to the ``bad
cavity'' case, where the damping rate $\gamma_{\rm cav}/\tau$ of the cavity is larger than
the one of the atomic polarisation $\Gamma $/2, and for which the
theoretically predicted squeezing is the best. The second mirror is a highly
reflecting one. The mode matching efficiency of the cavity with the input
beam is of the order of at least 96\% in power.
\begin{figure}
\centerline{\psfig{figure=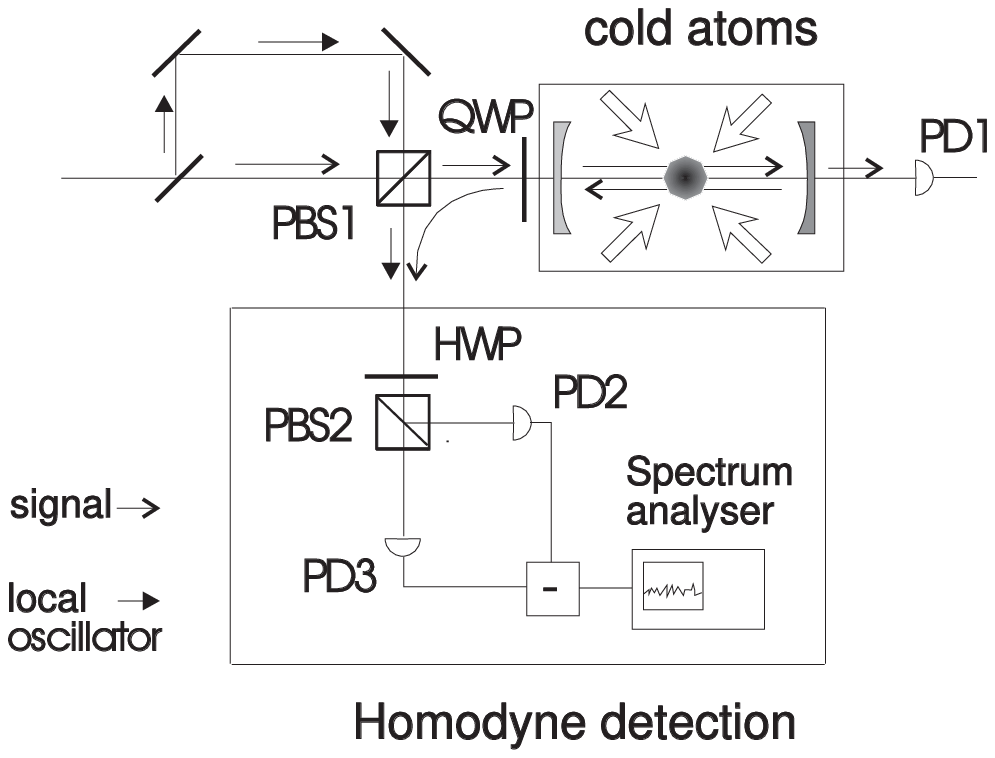,width=6.5cm}}
\caption{Experimental layout to measure the quantum noise of a probe beam
having passed through the cavity. PBS1, PBS2: polarising beamsplitters, QWP:
quarter-wave plate, HWP: half-wave plate, PD1, PD2, PD3: photodiodes; PD1
measures the transmitted power; PD2 and PD3 are employed in the homodyne
detection to record the reflected power and the quantum noise.}
\end{figure}

As already mentioned, the probe laser beam is generated by the same
Ti:Sapphire laser as the trapping beams. This laser has the main advantage
over diode lasers that it is free from any excess noise in the MHz range.
However, this imposes to adjust the frequency of the various beams with the
help of acousto-optic modulators. The only beam which is not
frequency-shifted is the probe beam, in order to avoid excess noise produced
by the AOM.

In the presence of the trap, the atom cloud is carefully centred on the
probe beam by maximising the emission of the ``cold atom Raman laser''~\cite
{Hilico92b}, which spontaneously oscillates even in the absence of a probe
beam when the cavity is well aligned around the cloud.

The probe beam coming out of the cavity is separated from the input beam by
an optical circulator, made of a polarising beamsplitter and a quarter-wave
plate, and mixed with a local oscillator, 100 times more intense than the
probe beam, using the second input port of the same beamsplitter. At the
output of the beamsplitter, the signal and the local oscillator beams are
perfectly aligned and have orthogonal polarisations. A half-wave plate and a
second polarising beamsplitter are used to split the combined beam of signal
and local oscillator into two beams of equal power, which are then detected
by two identical EG\&G FFD-100 photodiodes, operated with a bias voltage of
80V. The quantum efficiency of the photodiodes was found to be 86\%; it was
further increased to 94\% by the method of refocusing the reflected light.
The low frequency components of photocurrents are dc coupled to provide a
direct measurement of the mean power. The high frequency components are ac
coupled, amplified and subtracted to monitor the quantum noise of the probe
beam. The common mode rejection generally achieved in this balanced homodyne
detection is better than 20 dB in the range of 0 to 20MHz. The noise signal
is further amplified, sent to a spectrum analyser and recorded in a digital
storage oscilloscope. The mode matching between the signal and the local
oscillator is optimised by measuring the dc visibility of the interference
fringes on one detector when the phase of the local oscillator is scanned.
The efficiency of mode matching at resonance is usually found to be of the
order of 90\%. The total homodyne efficiency varies between 85\% at
resonance and 90\% on the sides of the resonance curve.

The fluctuations of the field are detected at a fixed analysis frequency 
$\Omega $/2$\pi $, while the phase of the local oscillator is rapidly varied
with a piezoelectric transducer and the length of the cavity is scanned
across resonance by the escape of the atoms after the trap is turned off.
When the cavity is close to resonance, the presence of the non-linear medium
manifests itself by a phase dependent noise.

Detection of squeezing during the rather narrow time slot in which the atoms
stay inside the beam is a non trivial problem. Having only
20-30 milliseconds to perform the measurement, important variations like the
phase scan of the local oscillator have to take place at frequencies of the
order of kHz. To record a spectrum, the bandwidth of the spectrum analyser
has consequently to be chosen equal to or larger than 100kHz. On the other
hand squeezing is quite sensitive to the observation frequency, thus a
bandwidth larger than 1MHz has to be avoided not to average out the effect.
A less obvious difficulty consists in the use of the so called videofilter,
which is usually a simple postdetection low-pass filter, acting only on the
display of the analysed signal to extract the essential information, which
would otherwise be covered by high frequency noise. Its bandwidth is usually
chosen to be 1/300 of the analysing bandwidth, i.e. 300Hz for a bandwidth
of 100kHz.

The signal to be analysed, that is the noise power of the light beam, is a
sine function having a fundamental frequency of the order of 1kHz. As a
consequence a linear (amplitude) or logarithmic (dB) display will not only
involve the fundamental frequency, but also higher harmonics. In particular
in Fourier space, an important part of information about the squeezing will
be stored in higher harmonics and the better the squeezing is the more
important higher harmonics are. Being displayed on a linear scale, a signal
containing up to 50\% squeezing in power, will contain no visible squeezing
anymore, if a postdetection video bandwidth of 300Hz is used. We thus had
to implement better adapted linear filters than a spectrum analyser can
supply, which means numerical filters.

The spectra should then be recorded without videofilter at all, converted
into noise power and filtered afterwards by a numerical RC filter at a well
adapted frequency to reduce the high frequency noise. As we know exactly the
frequency response of the videofilter, it is also possible to record a
filtered signal, reconstruct the original one numerically in the frequency
domain and to filter this signal converted into noise power with a numerical
low-pass at a better adapted frequency. The numerical filters were compared
to usual electrical RC filters; the effect on the spectrum was found to be
the same. They are thus equivalent to standard postdetection videofilters,
but have the essential advantage that their RC time can be freely chosen.

Figure 6 shows a spectrum reconstructed from a filtered one using the
procedure above described. It displays a recording of the noise power
obtained at a fixed geometrical cavity length while the changing number of
atoms is scanning the cavity across resonance. Out of resonance (on the left
hand side of the figure) the probe beam is totally reflected by the cavity
and is hence at shot noise. The average shot noise level is indicated by the
straight line. When the cavity becomes resonant with the probe beam, the
noise exhibits large oscillations due to the phase scan of the local
oscillator, which indicate the presence of phase dependent fluctuations. On
the lower branch of the bistability curve, i.e. before the system switches
towards high intensities, a quantum noise reduction of about 30\% within $%
\pm $ 10\% can be seen, whereas on the upper branch large excess noise is
observed in some quadratures. If one takes into account losses in
propagation, detection and matching, a value of the squeezing better by 4\%
can be inferred. This trace was obtained at an observation frequency of $%
\Omega $/2$\pi $=5MHz. The cavity linewidth ($\gamma _{{\rm cav}}$/2$\pi
\tau $=5MHz) was about as large as the atomic linewidth. Probe laser and
local oscillator were detuned by 10$\Gamma $ on the high frequency side of
the atomic resonance and their powers were 25$\mu$W and 9mW respectively.
The bistability parameter $C$ was found to be about 300 right after turning
off the trapping beams.

\begin{figure}[t]
\centerline{\psfig{figure=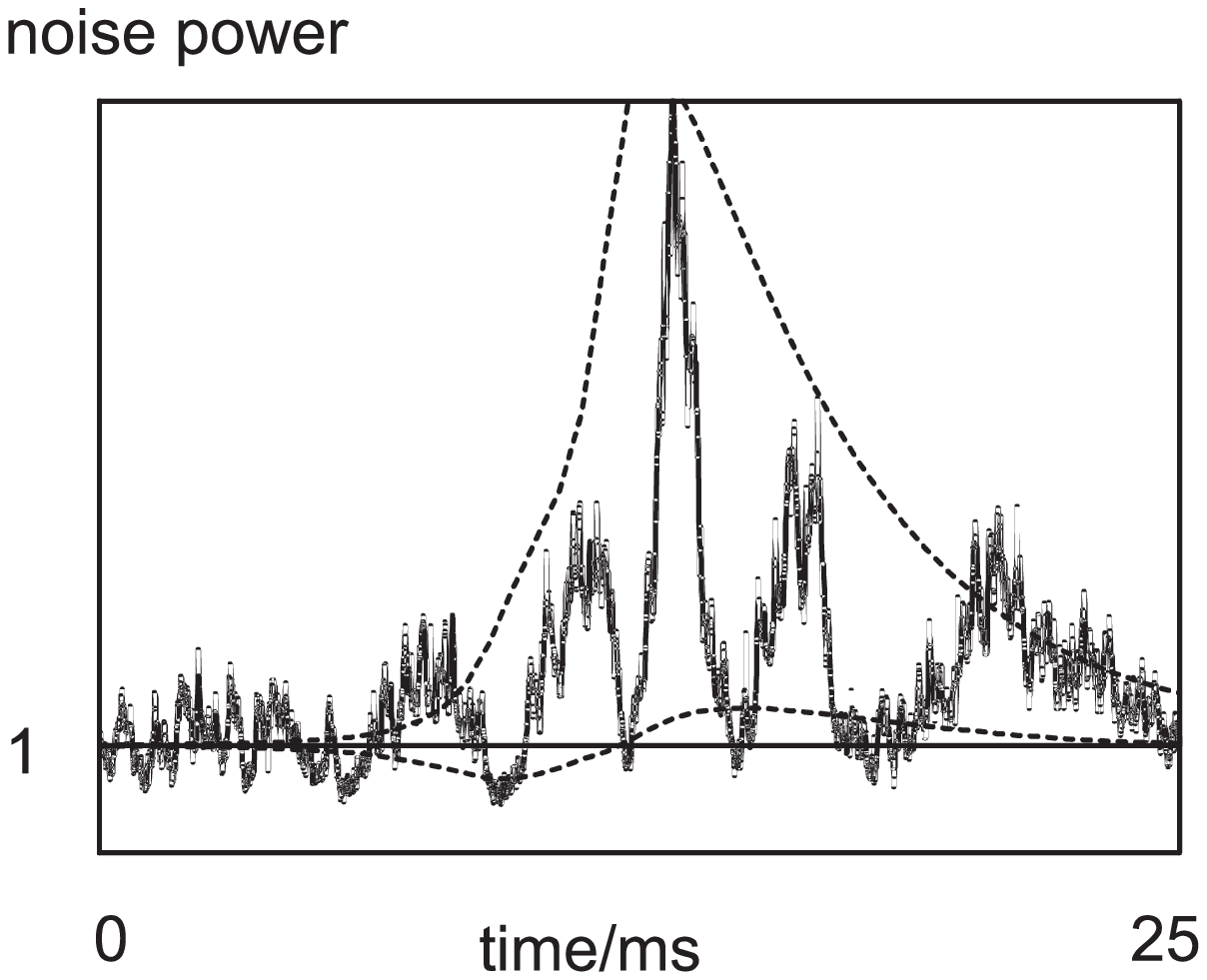,width=6.5cm}}
\caption{Recording of the quantum noise power at $\Omega/2\pi$=5MHz as a function of
time, while the decaying number of atoms is scanning the cavity across
resonance. The probe beam fluctuations drop below shot noise on the lower
branch of the bistability curve (left hand side of trace). Probe beam and
local oscillator were detuned by 10$\Gamma $ to the red side of the atomic
resonance. Their powers were 25$\mu$W and 9mW respectively. The straight
line indicates the average shot noise level; the broken lines correspond to
theoretical predictions of the minimal and maximal noise power using the
experimental data. The experimental trace was taken with a postdetection
video bandwidth of 300Hz, reconstructed numerically, converted into noise
power and low pass filtered at 1kHz afterwards.}
\end{figure}

This result has been compared with theoretical predictions given by the
two-level atom model derived from [12] including the experimental
parameters. This model was developed within the plane-wave approximation and
uses a linear input-output formalism to calculate the modifications of
quantum fluctuations of a coherent field after interaction with an optical
ring cavity containing an ensemble of two-level atoms. In agreement with the
experimental situation the cavity is scanned at a fixed geometrical length
by the change in the number of atoms interacting with the probe beam. In a
first approximation the relevant number of atoms is assumed to decay
exponentially in time. The quantum noise power in all quadratures can then
be calculated as a function of number of atoms inside the probe beam and
thus as a function of time. The theoretical curves shown as broken lines in
Figure 6 have been computed for the experimental data given above, supposing
an initial bistability parameter of 300. The two traces correspond to the
minimal and maximal noise power respectively and are in reasonable agreement
with the experiment. The next experimental step will consist in finding the
best conditions for squeezing by exploring carefully the parameter space. On
the theoretical side a more exact treatment of the interaction, including in
particular the temporal variation of the number of atoms and the effect of a
radially varying electric field, will be carried out.

\section{Conclusion}

Our study shows that cold atoms constitute a promising medium for non-linear
and quantum optics. The fact that it is possible to generate squeezed light
placing this medium inside an optical cavity opens the way to quantum optics
using cold atoms and hence provides an interesting alternative to atomic
beams. One of the difficulties of this experiment, coming from the fact that
the trapping beams have to be turned off during the measurements, should be
possible to circumvent in the future by working on a transition not coupled
to the trapping transition or by using a ``dark SPOT'' \cite{Ketterle93}%
{\large \ }demonstrated recently.

{\bf Acknowledgements:} The authors sincerely thank A. Heidmann for his very
valuable help. This work has been supported in part by the EC contract
ESPRIT BRA 6934 and the EC HMC contract CHRX-CT 930114.

\end{document}